\documentclass[a4paper,11pt]{article}
\usepackage{pos}
\usepackage{bm}
\usepackage{caption}
\usepackage[numbers,sort&compress]{natbib}
\usepackage{lipsum}

\let\OLDthebibliography\thebibliography
\renewcommand\thebibliography[1]{
  \OLDthebibliography{#1}
  \setlength{\parskip}{0pt}
  \setlength{\itemsep}{0pt plus 0.3ex}
}
\usepackage[capitalise]{cleveref}
\title{Thermalization of a jet wake in QCD kinetic theory}

\author*[a]{Fabian Zhou}
\author[b]{Jasmine Brewer}
\author[a]{Aleksas Mazeliauskas}

\affiliation[a]{Institut für Theoretische Physik, Universität Heidelberg,\\
  Philosophenweg 12, 69120 Heidelberg, Germany}

\affiliation[b]{Theoretical Physics Department, CERN,\\
 CH-1211 Genève 23, Switzerland}

\emailAdd{zhou@thphys.uni-heidelberg.de}

\abstract{We study the energy deposition of a high-momentum parton traveling through a Quark-Gluon Plasma using QCD kinetic theory. We show that the energy is first transported to the soft sector by collinear cascade and then isotropised by elastic scatterings. Remarkably, we find that the jet wake can be well described by a thermal distribution function with angle-dependent temperature. This could be used for effective phenomenological descriptions of jet thermalization in realistic heavy-ion collision simulations.}

\FullConference{HardProbes2023\\
 26-31 March 2023\\
 Aschaffenburg, Germany\\}

\begin{document}
\maketitle

\section{Introduction}
The observed suppression of high momentum particles in heavy ion collisions indicates the presence of a high density medium which the parton traverses. While interacting with its surroundings, the parton loses energy via different scattering processes. Hence this energy is deposited to the background. This phenomenon is referred to as \emph{jet quenching} \cite{Majumder:2010qh, Qin:2015srf} which allows to study the properties of the quark-gluon plasma (QGP).\\
A promising tool to describe the evolution of a jet inside an initially far-from-equilibrium plasma is an \emph{effective kinetic theory} (EKT) of QCD \cite{Arnold:2002zm}. EKT captures both, the hard and the soft sector of the dynamics under investigation, therefore it is suitable to describe the interplay between jet quenching and the thermalization of an expanding medium.

\section{Linearized effective kinetic theory}

The framework of EKT is a conformal leading order description in the coupling $\lambda = 4 \pi N_c \alpha_s$. We consider a one-particle distribution function $f(t,\bm{x}, \bm{p})$  whose dynamics is governed by the Boltzmann equation
\begin{equation}
    p^\mu \partial_\mu f(t,\bm{x}, \bm{p}) = -C[f].
\end{equation}
The collision kernel $C[f]$ contains all QCD inelastic $1\leftrightarrow 2$ and elastic $2\leftrightarrow 2$ processes. Once the heavy ion collision takes place, the system is rapidly expanding in longitudinal direction leading to an approximate boost invariance. Therefore it is convenient to use the proper time $\tau=\sqrt{t^2-z^2}$. Moreover, we assume the system to be spatially homogeneous and neglect transverse gradients.\\
During the evolution, most of the fireball consists of soft particles, whereas the occupancy of high momentum partons is small. This justifies a decomposition of the distribution function into a background and a linear perturbation $f = \Bar{f} + \delta f$, where a jet parton is described by $\delta f$. Under these assumptions the equations of motion become
\begin{align}
    \left(\partial_{\tau} - \frac{p_z}{\tau}\partial_{p_z}\right) \Bar{f}(\tau,\bm{p}) &= -C[\Bar{f}],\label{bg_equation}\\
    \left(\partial_{\tau} - \frac{p_z}{\tau}\partial_{p_z}\right) \delta f(\tau,\bm{p}) &= -\delta C[\Bar{f},\delta f],\label{pert_equation}
\end{align}
where the second term accounts for the expansion in beam direction. The background evolution in \cref{bg_equation} has been intensively studied in the past \cite{Kurkela:2015qoa, Kurkela:2018oqw}. In order to derive \cref{pert_equation} we assumed $\delta f \ll \Bar{f}$ \cite{Kurkela:2014tea,Kurkela:2018vqr,Mehtar-Tani:2022zwf}. Because it is a linear equation in $\delta f$, we can in practice solve it using an arbitrary normalization of $\delta f$. For the initial perturbation, we choose back-to-back jets,
\begin{equation}\label{init_g_pert}
    \delta f(\tau_0,\bm{p}) = J(\bm{p}) + J(-\bm{p}),
\end{equation}
with zero net momentum. The jet $J(\bm{p})$ represents an energetic particle with momentum $p = E$. Thus, we model the jet by a Gaussian in momentum space located at $\bm{p}_0 = (0,0,E)^T$
\begin{equation}
    J(\bm{p}) = \left(\frac{2\pi}{\sigma^2E^2}\right)^{3/2}\frac{E}{p} \exp\left(-\frac{p_x^2 + p_y^2 + (p_z - E)^2}{2\sigma^2E^2}\right),
\end{equation}
with $\sigma = 0.1$. For convenience we choose
\begin{equation}
    \int \frac{d^3p}{(2\pi)^3}p^0J(\bm{p}) = E.
\end{equation}
Close equilibrium, the main interest lies in the conserved quantities of the system such as energy and momentum. Therefore, it is useful to study the energy momentum tensor $T^{\mu\nu}$. It can be computed directly from the full distribution $f(\tau, \bm{p})$
\begin{equation}
    T^{\mu\nu}(\tau) = \int \frac{d^3p}{(2\pi)^3} \frac{p^\mu p^\nu}{p^0} f(\tau, \bm{p}).
\end{equation}
In a local rest frame, it takes the form
\begin{equation}
    T^{\mu\nu}(\tau) = \textrm{diag}(e, P_T, P_T, \frac{1}{\tau^2}P_L)^{\mu\nu},
\end{equation}
where $e$ is the energy density, $P_T$ is the transverse and and $P_L$ the longitudinal pressure.

\section{Angular dependent equilibration}

For asymptotically late times, the jet is expected to become part of the background. In case of a static one, we want to study how the jet is reaching thermal equilibrium. For an expanding background, we are addressing the question if the jet is \emph{hydrodynamizing}. The results are mostly obtained for a pure gluonic system.

\subsection{Non-expanding thermal background}

\begin{figure}
    \centering
        \includegraphics[height=0.3\textwidth]{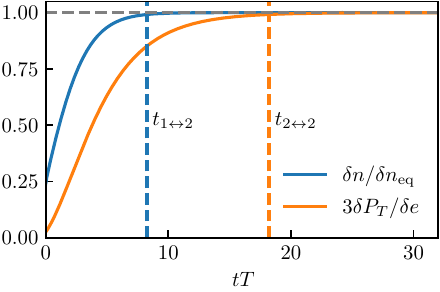}
        \includegraphics[height=0.3\textwidth]{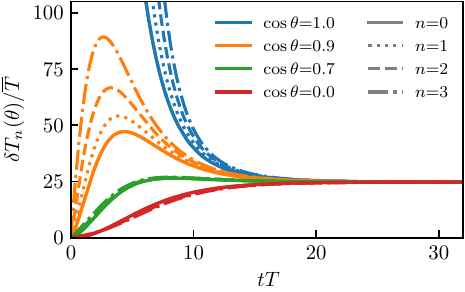}
    \caption{(left) Particle number $n$ and transverse pressure $P_T$ approaching there equilibrium value and (right) the collapse of the temperature perturbations $\delta T_n(\theta)$.}
    \label{fig:dIn_thermal_10_09_07_00}
\end{figure}
Firstly, we consider perturbations on top of a static background $\Bar{f} = f_{\textrm{eq}}$, where $f_{\textrm{eq}}$ is the Bose-Einstein distribution. Without expansion the time variable under consideration is $t$. The energy scale, at which the jet is set initially, is $E=15T$, where $T$ is the temperature of the background.\\
While the jet is evolving, the particle number $\delta n$ as well as the transverse pressure $\delta P_T$ is built up. While $\delta n$ is driven by the splittings, $\delta P_T$ encodes the momentum broadening coming from the elastic scatterings. As one can see in \cref{fig:dIn_thermal_10_09_07_00} (left), the equilibrium value of $\delta n$ is reached well before it is reached for $\delta P_T$, $t_{1\leftrightarrow 2} < t_{2\leftrightarrow 2}$. This shows that around $t \approx t_{1\leftrightarrow 2}$ the jet has already obtained some sort of equilibrium mainly originated from the splittings. This can further be seen using angular dependent moments of the distribution $\delta f$ defined by
\begin{equation}\label{moments}
    I_n(t, \theta) \equiv 4\pi\int \frac{p^2dp}{(2\pi)^3}p^n f(t, p,\theta) = \mathcal{N}_n T_n(t, \theta)^{n+3},
\end{equation}
where $\theta$ is the angle with respect to the $z$-axis, in this isotropic set up we choose the jet axis to be the same. The moments $I_n(t, \theta)$ provide an angular dependent effective temperature $T_n(t,\theta) = \overline{T}+\delta T_n(t,\theta)$. Importantly, in thermal equilibrium the temperature does not depend on $\theta$ (isotropy) and moreover $\mathcal{N}_n$ is chosen so that they agree for all moments $n$. With \cref{moments} we obtain an expression for the temperature perturbation $\delta T_n(t, \theta)$ that the jet induces once it equilibrates
\begin{equation}
    \frac{\delta T_n(t, \theta)}{\overline{T}} = \frac{\delta I_n(t, \theta)}{(n+3)\Bar{I}_n},
\end{equation}
where in this case the background quantities are isotropic. In \cref{fig:dIn_thermal_10_09_07_00} (right) we show the time evolution of $\delta T_n(t, \theta)/\overline{T}$\footnote{Because of our normalization of $\delta f$, $\delta T_n$ can be larger than $\overline{T}$.}. As soon as the curves collapse for different moments $n$, they are clearly still very different for different angles $\theta$. This indicates that the momentum distribution in each $\theta$-slice takes the form of a thermal distribution with temperature $T(\theta) = \overline{T} + \delta T(\theta)$.\\
Once the jet perturbation is fully in equilibrium, it takes the following form
\begin{equation}
    \delta f_{\textrm{eq}}(p) = \delta T \partial_T f_{\textrm{eq}}(p/T).
\end{equation}
Note that in the linearized theory the temperature perturbation $\delta T$ is the magnitude. So around the time $t \approx t_{1\leftrightarrow 2}$, when the distribution is thermal along each $\theta$-slice but still anisotropic in $\theta$, the jet is described by
\begin{equation}\label{unscaled_distr.}
    \delta f(p,\theta) \approx \delta T(\theta) \partial_T f_{\textrm{eq}}(p/T),
\end{equation}
\begin{figure}
    \centering
    \includegraphics[height=0.3\textwidth]{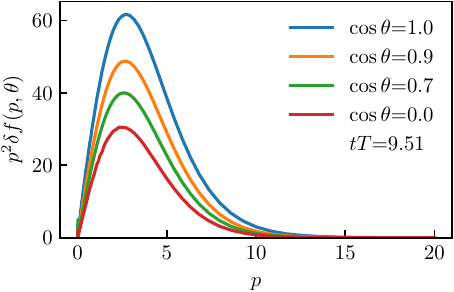}
    \includegraphics[height=0.3\textwidth]{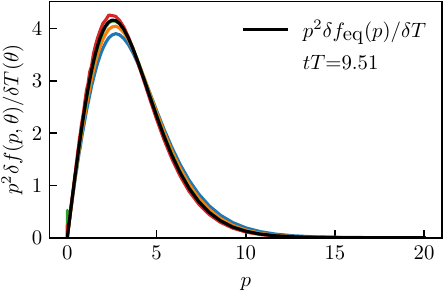}
    \caption{Momentum distributions of the jet at $t \approx t_{1\leftrightarrow 2}$ for different angles $\theta$, originally (left) and normalized to 1 (right).}
    \label{fig:unscaled_scaled}
\end{figure}
Therefore the shape is the same, only the magnitude $\delta T(\theta)$ is angular dependent. We demonstrate that by changing the normalization and demanding $\int dp p^2\delta f(p,\theta)=1$. This amounts to dividing out the amplitude in \cref{unscaled_distr.}. In \cref{fig:unscaled_scaled} we show both unscaled and scaled distributions for different angles $\theta$. While the system is still anisotropic, the scaled distributions $\delta f(p,\theta)/ \delta T(\theta)$ are in good agreement with the equilibrium distribution $\delta f_{\textrm{eq}}(p)/\delta T =  \partial_T f_{\textrm{eq}}(p/T)$.
\vspace{3mm}

\noindent
\begin{minipage}{.49\textwidth}
Adding fermions to the collision kernel $C[f]$ in \cref{bg_equation,pert_equation} we can study how the system equilibrates chemically. In that case we initialize the evolution with a gluon jet $\delta f_g(\tau_0,\bm{p})$ as in \cref{init_g_pert}, whereas the initial quark perturbation is set to zero, $\delta f_q(\tau_0,\bm{p})=0$. In \cref{fig:chem_eq}, we show how the equilibrium ratio of quark and gluon densities is approached. This happens after system reaches kinetic equilibrium, i.e. isotropization. Adding the expansion, the order of these equilibration processes is reversed (not shown).
\end{minipage}\quad
\begin{minipage}{.49\textwidth}
    \includegraphics[width=\linewidth]{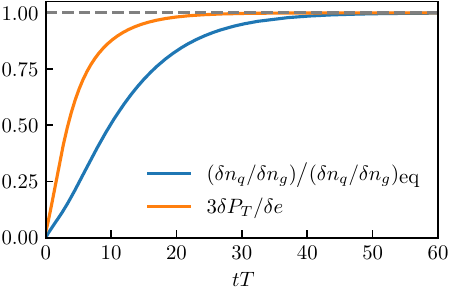}
    
    \vspace{-2mm}
    \captionof{figure}{Evolution of the quark to gluon number ratio and transverse pressure for a gluon jet.}
    \label{fig:chem_eq}
\end{minipage}

\subsection{Expanding non-equilibrium background}

Now we consider the background to be out-of-equilibrium and the system to be longitudinal expanding. It evolves in proper time $\tau$. The initial jet perturbation in \cref{init_g_pert} is now pointing in the transverse plane at $\bm{p}_0 = (E,0,0)^T$. Following \cite{Kurkela:2015qoa} we choose the initial condition for the background to be
\begin{equation}\label{init_prl_distr}
    \Bar{f} \left(\tau_0,p,\theta\right) = \frac{2AQ_0}{\lambda}\frac{e^{-\frac{2}{3}\frac{1}{Q_0^2}(p_T^2 + \xi^2 p_z^2)}}{\sqrt{p_T^2 + \xi^2 p_z^2}},
\end{equation}
where $\xi$ determines the anisotropy of the initial distribution, $Q_0$ is related to the saturation scale $Q_s$ and $A$ is constant that is fixed such that the comoving energy density $\tau \epsilon$ is in agreement with classical early time simulations. It is shown that the background hydrodynamizes, exhibiting memory loss about the initial conditions while doing so. We show how the jet is undergoing a similar process, approaching $\delta f_{\textrm{hydro}}$ (as opposed to $\delta f_{\textrm{eq}}$ in the thermal case). While there is an analytical formula for $\delta f_{\textrm{eq}}$ that makes it easier to compare with, there is no such expression for $\delta f_{\textrm{hydro}}$. For that reason we perturb the amplitude $A$ in \cref{init_prl_distr}. This gives us an \emph{azimuthally symmetric} perturbation
\begin{equation}
    \delta f_{\textrm{sym}}^{\textrm{az}}(\tau_0,p,\theta) = \frac{\delta A}{A}\Bar{f} \left(\tau_0,p,\theta\right),
\end{equation}
which is just the background multiplied by a small scale factor. By comparing the time evolution with the jet perturbation $\delta f$ we can study hydrodynamization. To this end, it is convenient to express the time in units of the relaxation time, leading to the scaled time variable $\Tilde{\omega} = \tau/\tau_R$ with $\tau_R = \frac{4\pi\eta/s}{T_{\textrm{eff}}}$. A crucial quantity to measure how far the perturbations have equilibrated is the normalized pressure anisotropy, defined by
\begin{equation}
    \mathcal{A}\equiv \frac{\delta P_T - \delta P_L}{\delta e/3},
\end{equation}
where the equilibrium pressure of a conformal system is given by $\delta P=\delta e/3$.\\
\begin{figure}
    \centering
    \includegraphics[height=0.3\textwidth]{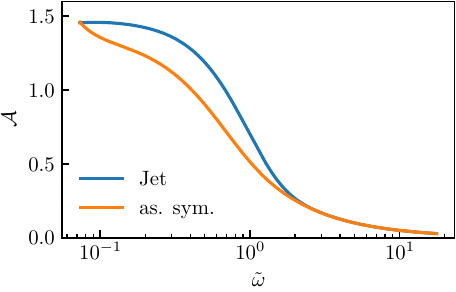}
    \includegraphics[height=0.3\textwidth]{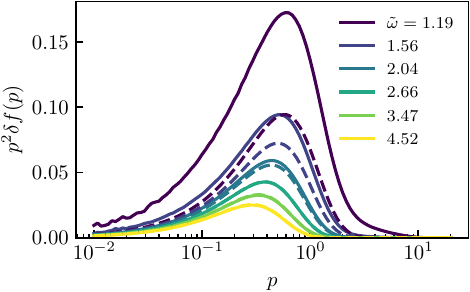}
    \caption{(left) Collapse of different perturbations for the normalized pressure anisotropy $\mathcal{A}$ and (right) the distributions. The solid line corresponds to $\delta f$ and the dotted line to $\delta f_{\textrm{sym}}^{\textrm{az}}$ at $\cos\theta = 0$, i.e. along the jet axis.}
    \label{fig:A_pert}
\end{figure}
In \cref{fig:A_pert} (left) we plot the $\mathcal{A}(\Tilde{\omega})$ as a function of scaled time for both, the jet perturbation as well as the azimuthally symmetric perturbation. The jet isotropizes much slower in the beginning, nevertheless, around $\Tilde{\omega}\approx 2$ both curves collapse into each other with $\mathcal{A}(\Tilde{\omega}\approx 2) \neq 0$. While it has been shown in the past that perturbations at the scale $Q_0$ hydrodynamize around $\Tilde{\omega}\approx 1$ \cite{Kurkela:2018vqr}, the jet takes more time due to its energy that is much higher than typical momentum scales of the background \cite{Kurkela:2014tea}. Furthermore, even the distributions $\delta f$ and $\delta f_{\textrm{sym}}^{\textrm{az}}$ themselves are indistinguishable around $\Tilde{\omega}\approx 2$ (\cref{fig:A_pert} (right)), similarly for larger $\cos\theta$ (not shown). This demonstrates a clear sign of hydrodynamization. 

\section{Conclusion}

We use EKT to describe the equilibration of a jet modelled by a linear perturbation on top of the background. Starting with a thermal non-expanding background, splittings already achieve that the distribution of the jet becomes thermal characterized by an angular dependent temperature. Only then the elastic scatterings isotropize the system. Including fermions, chemical equilibrium is reached after the system becomes isotropic. For an expanding non-equilibrium background, we observe the hydrodynamization of the jet which is delayed compared to this process involving only the background. This can be attributed to the higher initial momentum scale of the jet. Following the same procedure as in \cite{Kurkela:2018vqr}, one could compute the linear response to the initial jet perturbation in order to obtain a macroscopic description of the equilibration of the jet. This could be useful for phenomenological studies of the energy loss of jet inside an expanding medium.

\end{document}